\documentclass[twocolumn,aps,showpacs,prb]{revtex4}
\usepackage{graphicx}

\begin{document}
\title{The chemistry of La on the Si(001) surface from first
principles } 
\author{Christopher R. Ashman,$^{1}$ Clemens J. F\"orst,$^{1,2}$
Karlheinz Schwarz$^{2}$ and Peter E. Bl\"ochl,$^{1,*}$}
\affiliation{$^1$ Clausthal University of Technology, Institute for
Theoretical Physics, Leibnizstr.10, D-38678 Clausthal-Zellerfeld,
Germany}
\affiliation{$^2$ Vienna University of Technology, Institute for
Materials Chemistry, Getreidemarkt 9/165-TC, A-1060 Vienna, Austria}
\date{\today} 
\begin{abstract} This paper reports state-of-the-art electronic
structure calculations of La adsorption on the Si(001) surface. We
predict La chains in the low coverage limit, which condense in a stable
phase at a coverage of $\frac{1}{5}$~monolayer.  At $\frac{1}{3}$
monolayer we predict a chemically rather inert, stable phase.  La
changes its oxidation state from La$^{3+}$ at lower coverages to
La$^{2+}$ at coverages beyond $\frac{1}{3}$ monolayer. In the latter oxidation
state, one electron resides in a state with a considerable contribution
from La-$d$ and $f$ states.  \end{abstract}
\maketitle

\section{Introduction}

Device scaling has been the engine driving the microelectronics
revolution as predicted by Moore's law.\cite{Moore95} By reducing the
size of transistors, processors become faster and more power efficient
at an exponential rate. Currently the main challenge in device
scaling is the integration of high-k oxides as gate oxides into
silicon technology. 

The gate oxide is an integral part of a metal-oxide-semiconductor
field-effect transistor (MOSFET). It is the dielectric of a capacitor,
which is used to attract charge carriers into the channel between source
and drain, and thus switches the transistor between its conducting and
its non-conducting state.  With a thickness of approximately 1-2
nm,\cite{roadmap} the gate oxide is the smallest structure of a
transistor. Further scaling would result in an unacceptably high quantum
mechanical leakage current and thus a large power consumption. 

In current transistors, the gate oxide is made from SiO$_2$ and
SiO$_x$N$_y$.  Future transistor generations will have to employ oxides
with a higher dielectric constant (high-k). This allows greater physical
thicknesses and thus reduces the quantum mechanical leakage currents.
The main contenders for the replacement of SiO$_2$ in future transistors
are, from today's point of view, oxides containing alkaline earth metals
like Sr or Ba, third-row elements like Y or La, forth-row elements like
Ti, Zr and Hf, or mixtures thereof.  Prominent examples are perovskite
structures around SrTiO$_3$\cite{McKee98} and
LaAlO$_3$\cite{Cabanas97,Nieminen01,Park01}, fluorite structures like
ZrO$_2$ and HfO$_2$\cite{Wilk01} and also Y$_2$O$_3$ and
La$_2$O$_3$\cite{Guha00,Guha02} or pyrochlore structures like
La$_2$Hf$_2$O$_7$\cite{apostolopoulos04} and
La$_2$Zr$_2$O$_7$.\cite{Fompeyrine02,Seo03} Recently, also promising
results on Pr$_2$O$_3$ have been published.\cite{Osten01}

While the first high-k-oxides will be grown with an interfacial SiO$_2$
layer, a further reduction in scale requires high-k-oxides with a direct
interface to silicon.  The requirement to limit interface states, and
the often crystalline nature of the oxides demand an epitaxial growth of
the oxides on silicon. Considering layer-by-layer growth by molecular
beam epitaxy (MBE), the first growth step for high-k oxides is the
deposition of the metal on silicon. Therefore we have investigated
deposition of metals out of the three most relevant classes for high-k
oxides on Si(001). These are the divalent alkaline-earth metals and the
three- and the four-valent transition metals.  The results on adsorption
of Zr and Sr have been published previously.\cite{Foerst03,Ashman04} The
present paper completes the study with a description of La-adsorption on
Si(001) as example of a trivalent metal.

Our previous work has shown that Zr tends to form silicides
readily.\cite{Foerst03} Silicide grains have been observed after Zr
sputtering on Si(001),\cite{Sun00} unless silicide formation is
suppressed by early oxidation which, however, leads to interfacial
SiO$_2$.  The Sr silicides are less stable in contact with silicon and
due to their sizable mismatch in lattice constant, nucleation does not
proceed easily.  The alkaline-earth metals Sr and Ba have been used in
the first demonstration of an atomically defined interface between a
high-k oxide, namely SrTiO$_3$ and silicon.\cite{McKee98}  By following
through the detailed steps of the formation of this interface, starting
at the low-coverage structures of metal adsorption, we were able to
provide a new picture for the phase diagram of Sr on
Si(001).\cite{Ashman04}  The phase diagram has been important to link
the theoretical interface structure of SrTiO$_3$ on Si(001) to the
experimental growth process.\cite{Ashman04,Foerst04,Norga03} From the interface
structure and its chemistry we could show in turn that the band-offset,
a critical parameter for a transistor, can be engineered to match
technological requirements by carefully controlling the oxidation of the
interface.\cite{Foerst04}

Since many of the characteristics of Sr adsorption carry over to
La-adsorption let us briefly summarize the main results.\cite{Ashman04}
Sr donates its electrons to the empty dangling bonds of the Si-surface.
The Si-dimers receive electron pairs one-by-one, and unbuckle as they
become charged.  When all Si dangling bonds are filled, i.e. beyond
$\frac{1}{2}$~monolayer (ML), additional electrons enter the anti-bonding
states of the Si-dimers at the surface, and thus break up the
Si-dimer-row reconstruction.  

At low coverage, Sr forms chains running at an angle of 63$^\circ$ to the
Si-dimer rows. As the coverage increases, the chains condense first into
structures at $\frac{1}{6}$~ML and at $\frac{1}{4}$~ML, which are
determined by the buckling of the Si-dimers and their electrostatic
interaction with the positive Sr ions. At $\frac{1}{2}$~ML a chemically
fairly inert layer forms, where all dangling bonds are filled and all
ideal adsorption sites in the valley between the Si-dimer rows are
occupied.

The paper is organized similar to our previous work on Sr adsorption. In
Sec.~\ref{sec:compdet} we describe the computational details of the
calculation. In Sec.~\ref{sec:sisurf} and~\ref{sec:silicides} we review
briefly the reconstruction of the Si(001) surface and we discuss the
known bulk La silicides. Sec.~\ref{sec:iso+dimer}, \ref{sec:chains}
and~\ref{sec:condchains} deal with the low coverage limit, where La
ad-atoms form dimers and chain structures. Beyond the canonical coverage
of 1/3~ML (Sec.~\ref{sec:third}) we observe a change in the oxidation
state of the La ad-atoms from 3+ to 2+ (Sec.~\ref{sec:2+3+}).  The
results are placed into context in Sec.~\ref{sec:phasediagram} where we
propose a phase diagram for La on the surface.  The computational
supercells used for the simulation of the low-coverage structures are
shown in the appendix.

\section{Computational Details} \label{sec:compdet}
The calculations are based on density functional theory\cite{Kohn,
KohnSham} using a gradient corrected functional.\cite{PBE}  The
electronic structure problem was solved with the projector augmented
wave (PAW) method,\cite{PAW94,Blo03} an all-electron electronic
structure method using a basis set of plane waves augmented with partial
waves that incorporate the correct nodal structure. The frozen core
states were imported from the isolated atom.  For the silicon atoms we
used a set with two projector functions per angular momentum for $s$ and
$p$-character and one projector per angular momentum with $d$-character.
The hydrogen atoms saturating the back surface had only one $s$-type
projector function.  For lanthanum we treated the 5$s$ and 5$p$ core
shells as valence electrons.  We used two projector functions per
magnetic quantum number for the $s$, $p$, and $f$ angular momentum
channels and one for the $d$ channel.  The augmentation charge density
has been expanded in spherical harmonics up to $\ell=2$. The kinetic
energy cutoff for the plane wave part of the wave functions was set to
30~Ry and that for the electron density to 60~Ry.

A slab of five silicon layers was used as silicon substrate. This
thickness was found to be sufficient in previous studies on Sr
adsorption.\cite{Ashman04} The dangling bonds of the unreconstructed
back surface of the slab have been saturated by hydrogen atoms. The
lateral lattice constant was chosen as the experimental lattice constant
$a=5.4307~$\AA\ of silicon,\cite{CRC} which is 1~\% smaller than the
theoretical lattice constant.  Since we always report energies of
adsorbate structures relative to the energy of a slab of the clean
silicon surface, the lateral strain due to the use of the experimental
lattice constant cancels out.  The slabs repeat every 16~\AA\
perpendicular to the surface, which results in a vacuum region of
9.5~\AA\ for the clean silicon surface.

The Car-Parrinello ab-initio molecular dynamics\cite{Car85} scheme with
damped motion was used to optimize the electronic and atomic structures.
All structures were fully relaxed without symmetry constraints.  The
atomic positions of the backplane of the slab and the terminating
hydrogen atoms were frozen.

Many of the adsorption structures are metallic, which requires a
sufficiently fine grid in k-space. We used an equivalent to twelve by
twelve points per $(1\times 1)$ surface unit cell.  Previous
studies have shown that a mesh of eight by eight k-points is
sufficient.\cite{Ashman04} We have chosen a higher density here  as
this allows us to use commensurate k-meshes for $3\times$ and $2\times$
surface reconstructions.

For metallic systems, the orbital occupations were determined using the
Mermin functional,\cite{Mermin} which produces a Fermi-distribution for
the electrons in its ground state. The electron temperature was set to
1000~K. In our case this temperature should not be considered as a
physical temperature but rather  as a broadening scheme for the states
obtained with a discrete set of k-points.  The Mermin functional adds an
entropic term to the total energy, which is approximately canceled by
taking the mean of the total energy $U(T)$ and the Mermin-free energy
$F(T)=U(T)-TS(T)$ as proposed by Gillan:\cite{Gillan89}

\begin{equation}
U(T=0)\approx\frac{1}{2}(F(T)+U(T)).
\end{equation}

The forces are, however, derived from the free-energy $F(T)$. The
resulting error has been discussed previously.\cite{Ashman04}

In order to express our energies in a comprehensible manner, we report
all energies relative to a set of reference energies.  This set is
defined by bulk silicon and the lowest energy silicide LaSi$_2$. The
reference energies are listed in Tab.~\ref{tab:reference}. The reference
energy $E_0[\mathrm{La}]$ for a La atom, corresponding to the coexistence of bulk
silicon and bulk La, is extracted from the energy $E[\mathrm{LaSi_2}]$ of
the disilicide calculated with a $9\times9\times3$  k-mesh for the
tetragonal unitcell with $a=4.326$ and $c=13.840$ and the
reference energy of bulk silicon $E_0[\mathrm{Si}]$ as

\begin{equation}
E_0[\mathrm{La}]=E[\mathrm{LaSi}_2]-2E_0[\mathrm{Si}].
\end{equation}

The bulk calculation for silicon was performed in the two atom
unit cell with a ($10\times 10\times 10$) k-mesh and at the experimental
lattice constant of 5.4307~\AA.\cite{CRC} 

\begin{table}
\caption{Reference energies used in this paper without frozen core energy. See text for
details of the calculation.}
\label{tab:reference}

\begin{tabular}{lr}
\hline
\hline
                   & \hspace*{0.3cm} Energy [H] \\ \hline
$E_0$[Si]          &       -4.0036\\
$E_0$[La]          &      -32.1395 \\
$E_0^{(1\times 1)}$[5 layer-Si-slab] &  -21.1139 \\
\hline
\hline
\end{tabular}
\end{table}

For the surface calculations, we always subtracted the energy of a slab
of the clean ($4\times2$) silicon surface of the same layer thickness
and backplane.

\section{The silicon surface}
\label{sec:sisurf}

Before discussing the adsorption of La, let us briefly summarize the
chemistry of the clean (001) surface of silicon. A more detailed
account has been given previously.\cite{Ashman04}

On the unreconstructed silicon surface, the atoms form a square array.  Due to
a lack of upper bonding partners, each atom has two singly occupied dangling
bonds pointing out of the surface.  In order to avoid half-occupied bands,
pairs of silicon atoms dimerize, using up one dangling bond per atom to form
the dimer bond.  This is called the dimer row reconstruction.  Still, one
dangling bond per silicon atom is half occupied, which drives the so-called
buckled-dimer reconstruction: One atom of each dimer lifts up and the other
shifts down, resulting in a ``buckled'' dimer. This buckling causes both
electrons to localize in the upper, $sp^3$-like silicon atom of a dimer,
whereas the other, $sp^2$-like silicon atom with the empty p-like dangling
bond prefers a more planar arrangement.

\begin{figure}
\includegraphics[angle=0,width=8cm,clip=true,draft=false]{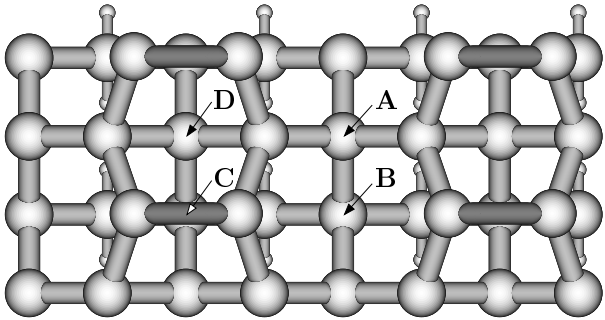}
\caption{Top view of the Si(001) surface and the four high
symmetry positions spanning the surface irreducible $(2\times1)$ unit
cell.  The dimer buckling is not shown.}
\label{fig:sisurf}
\end{figure}

Fig.~\ref{fig:sisurf} shows a ball stick model of the silicon surface
and introduces the high-symmetry adsorption sites A to D, which we will
refer to in the following discussion.

Filling the empty dangling bond with two electrons results in a removal
of the buckling as observed in our studies on Sr
adsorption.\cite{Ashman04} As La has an odd number of valence electrons,
we also examined the changes of the buckling upon filling the initially
empty dangling bond with a single electron. While the difference in $z$
coordinate of the two silicon atoms of a buckled dimer is 0.76\,\AA\ and
the one of an unbuckled dimer 0.00\,\AA, it is 0.35\,\AA\ after donation
of a single electron to a dimer. Thus the amplitude of the dimer
buckling may be used as a measure for the electron count.

\section{Bulk La silicides}
\label{sec:silicides}

In the case of Sr, the chemical interaction with silicon could be
understood by investigating the bulk Sr silicides.\cite{Ashman04} All
these structures could be explained by the Zintl-Klemm
concept.\cite{Zintl39} The electropositive Sr atoms donate their two
valence electrons to the silicon atoms. Each accepted electron saturates
one of silicon's four valences. A Si$^{-}$ can thus form three
covalent bonds, a Si$^{2-}$ only two, a Si$^{3-}$ only one and a
Si$^{4-}$ has a closed shell and does not form covalent bonds. This
principle was found to also be valid for the surface reconstructions of
Sr on Si(001).\cite{Ashman04}

The Zr silicides on the other hand cannot be explained by the simple
Zintl-Klemm concept.\cite{Foerst03} The Zr $d$ states also contribute to
the bonding and thus retain a variable number of electrons.

Similarly, the La silicides cannot be simply explained by a quasi-ionic
interaction with silicon. We find La in formal charge states between two
and three (i.e. charge according to the Zintl-Klemm concept). Also the
atom and angular momentum resolved density of states reveals, that La
$d$ states are partly occupied in these structures.  

\begin{figure}
\includegraphics[angle=0,width=8cm,draft=false,clip=true]{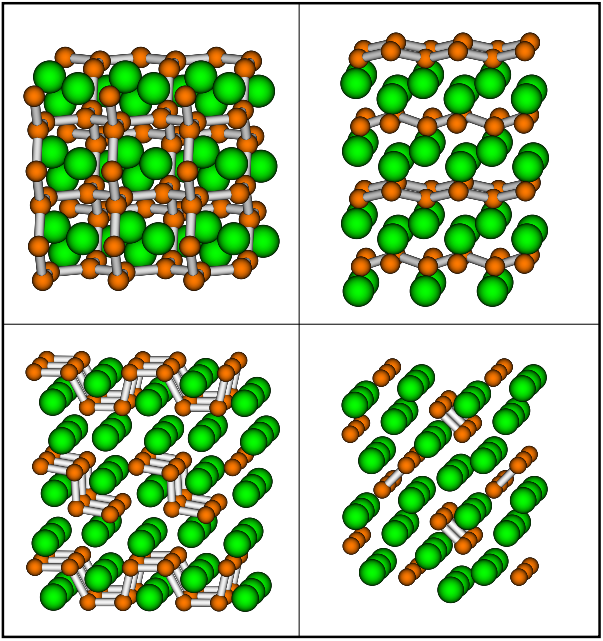}
\caption{(Color online) The bulk silicide structures. 
top left: LaSi$_2$ $I41/amdS$\cite{lasi2} (lowest energy structure); 
top right: LaSi $Pnma$;\cite{lasi_pnma} 
bottom left: LaSi $Cmcm$;\cite{lasi_cmcm} 
bottom right: La$_3$Si$_2$ $P4/mbm$.\cite{la3si2}}
\label{fig:lasilicides}
\end{figure}

Fig.~\ref{fig:lasilicides} shows the La silicides. LaSi$_2$ is the
lowest energy silicide. The energies per La atom are listed in
Table~\ref{tab:enesilicides}.

\begin{table}
\caption{Energies per La atom of bulk silicides relative to our
reference energies (Tab.~\ref{tab:reference}).}
\label{tab:enesilicides}

\begin{tabular}{lr}
\hline
\hline
                    &\hspace*{1cm} E[La] [eV]\\
\hline
LaSi$_2$ $(I41/amdS)$     &  0.00 \\
LaSi $(Pnma)$             &  0.42 \\
LaSi $(Cmcm)$             &  0.62 \\
La$_3$Si$_2$  $(P4/mbm)$  &  0.80 \\
\hline
\hline
\end{tabular}
\end{table}

\section{Ad-atoms and La dimers}
\label{sec:iso+dimer}

Our search for the adsorption structures of La have been guided by the
electron count rules that emerged from our investigation of
Sr\cite{Ashman04} adsorption on the same surface.  The studies of Sr
provided a consistent picture: The electrons from Sr are fully
transfered into the Si-dimer dangling bonds of the Si substrate.  The
ordering of Sr atoms on the surface is determined by the electrostatic
attraction between the Sr-cations and negatively charged Si-ions at the
surface. The negative Si-ions are the raised atoms of buckled Si dimers
and the atoms of filled, and thus unbuckled, dimers.  This picture holds
up to coverages where all Si-dimers are filled at 1/2~ML.  Due to the
different electron count of La as compared to Sr, we expect that the
silicon dimers are filled already at a coverage of 1/3~ML and secondly
we anticipate deviations from the above scheme.

Even though we predict La-chains to be the most stable structures in the
low coverage limit, we first investigate isolated
ad-atoms\cite{spin-pol} and La-dimers in order to provide an
understanding of the constituents of the more extended structures. Chain
structures are more stable by 0.34-0.40~eV per La atom as compared to
isolated ad-atoms.

Similar to Sr,\cite{Ashman04} we find the most stable position of an
isolated La atom at position $A$, in the center of four Si-dimers
(compare Fig.~\ref{fig:sisurf}). The position D, B and C have energies 0.23~eV,
0.51~eV and 1.70~eV higher than position $A$. A
$(4\times 4)$ supercell has been used for these calculations.

The diffusion barrier along the valley is equal to the energy difference
between sites $A$ and $B$, namely 0.52~eV, the one across the row is
1.31~eV and is estimated by the midpoint between the sites $A$ and $D$.

\begin{figure}
\includegraphics[angle=0,width=8cm,draft=false,clip=true]{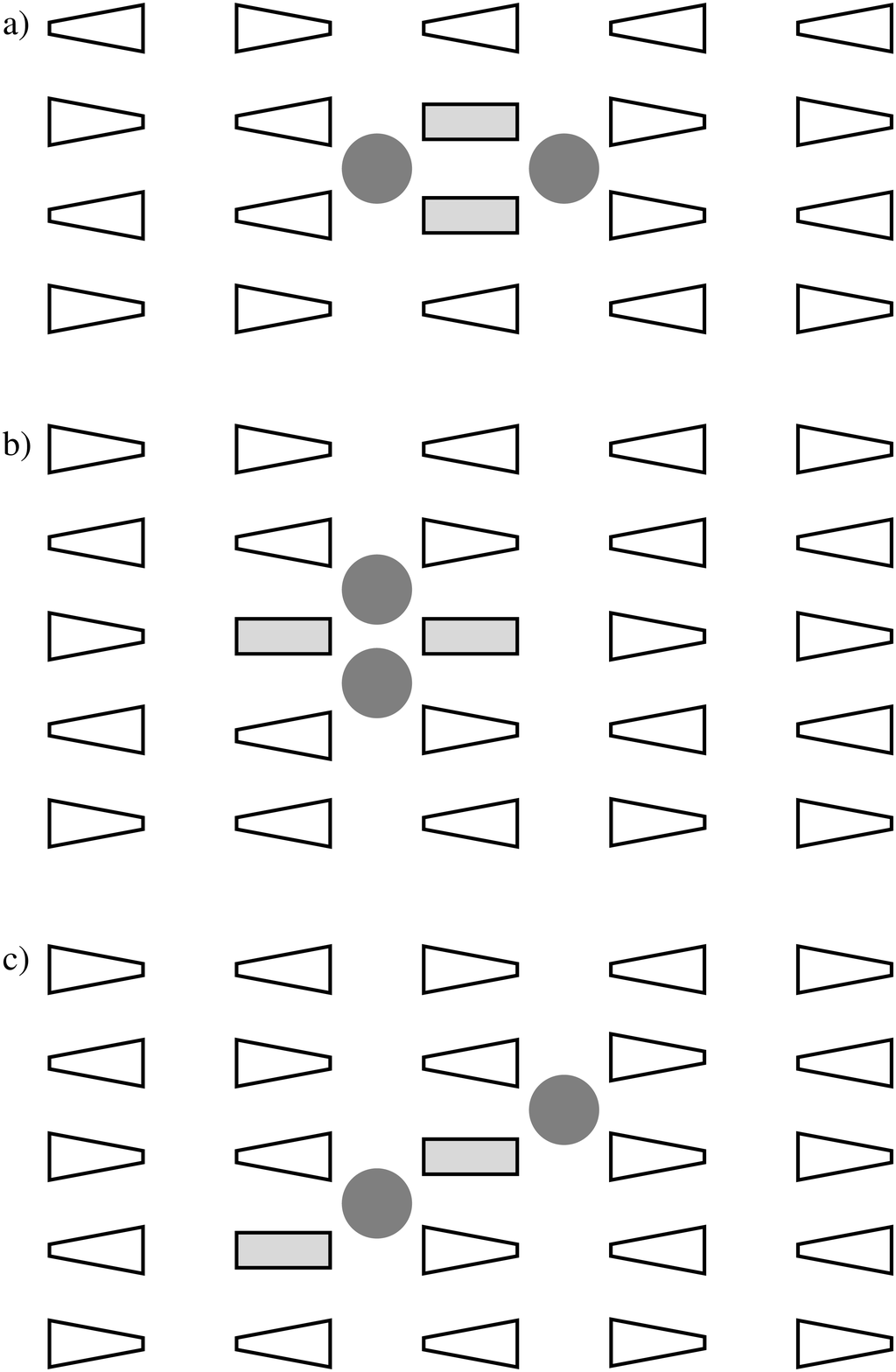}
\caption {Schematic representation of isolated La-dimers on the Si(001) surface.
The filled circles represent the La ad-atoms, the rectangles represent a
filled and therefore unbuckled Si-dimers. The triangles represent
buckled Si-dimers.  The flat side of a buckled dimer indicates the upper Si
atom with a filled dangling bond, whereas the pointed side indicates the
lower Si atom with the empty dangling bond. Only the Si-dimers
which are clearly unbuckled have been drawn as rectangles. The partially
unbuckled ones are represented by triangles (see discussion in
section~\ref{sec:iso+dimer}). The supercells used for the total energy
calculations are shown in Fig.~\ref{fig:dimersupercells}.
}
\label{fig:dimerorientation} 
\end{figure}

The formation of La dimers lowers the energy per ad-atom by
0.10--0.18~eV compared to isolated ad-atoms.  Due to the topology of the
Si(001) surface, three different types of La-dimers can be formed: (1)
orthogonal to the Si dimer rows, (2) parallel to the Si dimer rows and
(3) diagonal to the Si dimer rows.  All three structures are shown in
Fig.~\ref{fig:dimerorientation}. We find that the parallel La-dimer is
lowest in energy, followed by the orthogonal and diagonal La-dimers. All
La-dimers lie within a small energy window of 0.08\,eV.  Note, that we
only investigated singlet states.

A pair of La atoms has six valence electrons and from the lessons
learned from Sr adsorption, one would assume that three Si-dimers in the
vicinity of the La-dimer get unbuckled.  This is, however, not the case.
Only two Si-dimers become fully unbuckled. The remaining two electrons
from the La-dimer enter into states that are derived from the upper
dangling-bond band and which have an admixture of the La-$d$ and $f$
states. 

Usually one can easily distinguish between buckled and unbuckled dimers.
In the vicinity of La-dimers oriented diagonally or orthogonally,
however, we also observe Si-dimers with an intermediate buckling
amplitude.  Thus, in these cases, the oxidation state of the La atom,
namely 2+ versus 3+ cannot be attributed in a unique manner.

For the orthogonal and diagonal La-dimers we observe a tendency for the
La atoms to reduce their distance compared to staying centered on A
sites by 1~--~4\,\%.  For the parallel La-dimer this effect is opposite
and much larger.  The distance between the two La atoms is 4.11\,\AA ,
compared to 3.84\,\AA\ between two A sites which amounts to an expansion
of 7\,\%. This ad-atom repulsion within one valley has already been
observed in case of Sr\cite{Ashman04} and explains the formation of
isolated chains instead of condensed chains or clusters at low
coverages.  Nevertheless, we find the parallel dimer to be the most
stable.

\section{Chain structures at low coverages}
\label{sec:chains}

As we combined pairs of the most favorable site for isolated La ad-atoms
into dimers, we now search for ways to stack the three types of
La-dimers together into more extended structures.  

\begin{figure}
\includegraphics[angle=0,width=8cm,draft=false,clip=true]{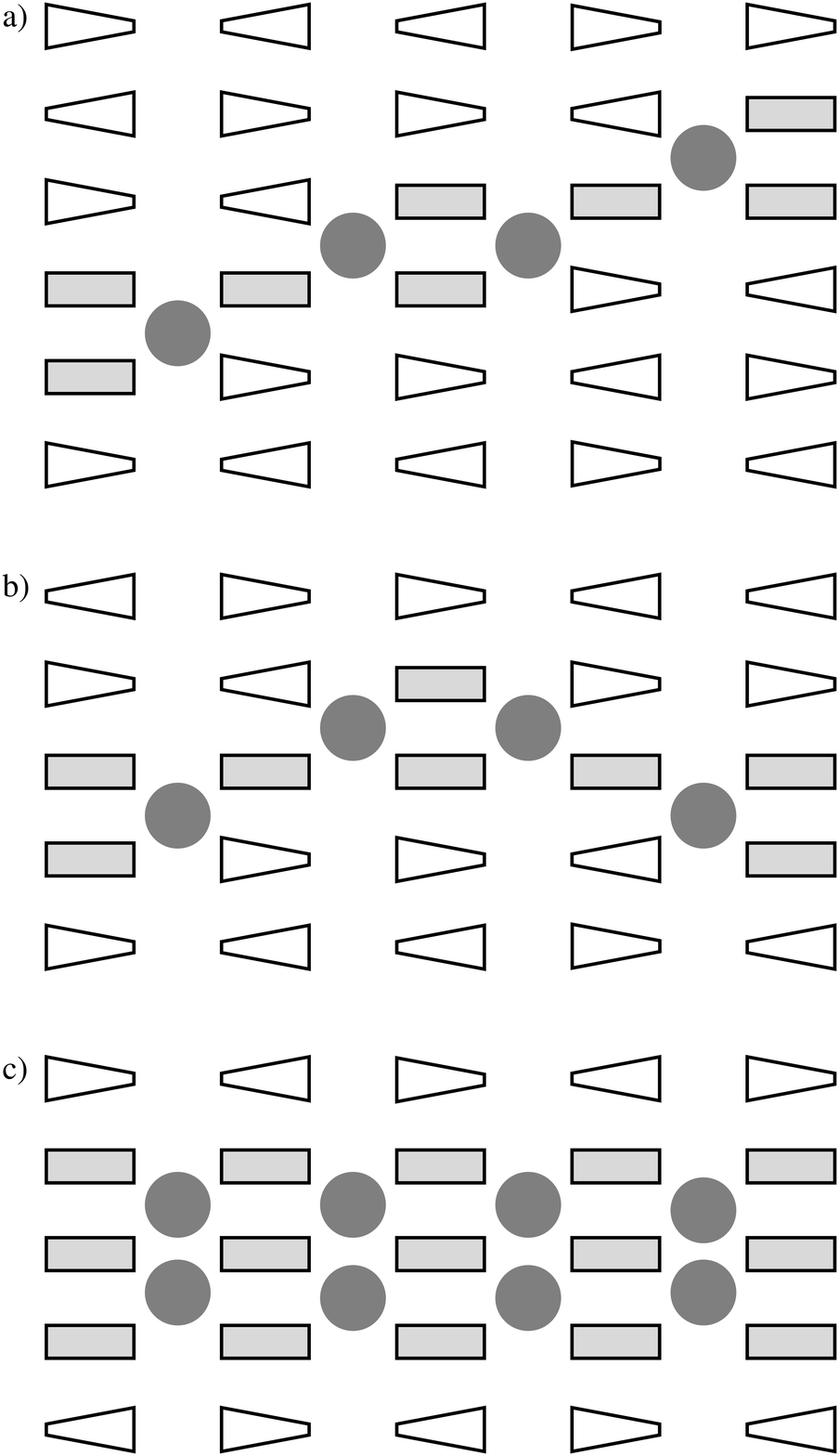}
\caption {Schematic representation of isolated La ad-atom chains.
a) a single, double-stepped La chain.
This is the energetically most favorable surface reconstruction at low
coverages. A change in chain direction is realized by stacking two
La-dimers with different orientation (b). Panel (c) shows the lowest
energy chain structure derived from parallel dimers.} 
\label{fig:lachain} 
\end{figure}

We systematically approached linear chain structures. Each of the three
La-dimer types -- parallel to the Si-dimer row, orthogonal or diagonal
-- has been stacked such that it shares at least one Si-dimer, so that
this Si-dimer is next to two Lanthanum atoms from different La-dimers.
The energetic ordering has been deduced on the basis of binding energy
per La atom (compare Tab.~\ref{tab:liniearchainenergies}). Note
that the binding energy per La atom for a given chain structure is
slightly coverage dependent. In case of the double stepped chains
(compare Fig.~\ref{fig:lachain}a for the structural principle) the
adsorption energy varies within 0.06\,eV at coverages between 1/10 and
1/5~ML. In order to ensure comparability all numbers listed here refer
to a coverage of 1/6~ML.

We start our investigation with parallel La-dimers shown in
Fig.~\ref{fig:dimerorientation}b, which is the most stable dimer
structure. The most favorable chain in this class is stacked
perpendicular to the Si-dimer rows as shown in Fig.~\ref{fig:lachain}c.
Its energy lies 0.06~eV per La atom above the lowest energy chain
structure.

The most favorable chain made from orthogonal La-dimers is shown in
Fig.~\ref{fig:lachain}a. It can also been interpreted as a variant of a chain
of diagonal La-dimers (compare Fig.~\ref{fig:dimerorientation} c). This is
the most favorable chain structure of La atoms on Si(001). Its chains run
at an angle of about 76$^\circ$ to the Si-dimer row. It should be noted
that it is equally possible to arrange the La-dimers in a zig-zag manner as
shown in ~\ref{fig:lachain}b. The zig-zag chain has not been explicitly
calculated.  The coexistence of straight and zig-zag chains has been
found for Sr on Si(001), where the two modifications have been shown to
be almost degenerate in energy.\cite{Ashman04}

\begin{table}
\begin{tabular}{lrrc}
\hline\hline
La-dimer type & $\alpha$ & E[La] [eV] & panel\\ \hline
parallel            & 90     & -0.30 & a \\
parallel            & 63           & -0.15 & b \\
parallel            & 45           & -0.20 & c \\
parallel      & 34     & -0.13 & d \\
parallel      & 0      & -0.07 & e \\
orthogonal          & 90      & -0.26 & f \\
orthogonal/diagonal\hspace*{0.5cm} & 76     & -0.36 & g\\
diagonal            & 63     & -0.28 & h\\
\hline \hline
\end{tabular}
\caption{Energies per La atom of the chain structures at 1/6~ML. The orientation
of the chain is described by the angle $\alpha$ (degrees) of the chain 
to the Si-dimer row. The supercells used for the total 
energy calculations are sketched in the
corresponding panels of Fig.~\ref{fig:supercells}.}
\label{tab:liniearchainenergies}
\end{table}

In all low-energy structures each La atom is thus surrounded by four
silicon atoms having filled dangling bonds.  Three of them are partners
of filled Si-dimers while one is a buckled Si-dimer with the negative Si
atom pointing towards the La ad-atom.  On the basis of counting
unbuckled Si-dimers, these structures are in a 3+ oxidation state.

The La-chain is the configuration with lowest energy in the low coverage
limit.  The lowest energy chain structures are of the order of 0.17~eV
per La atom more stable than the most favorable isolated La-dimer.  At
elevated temperatures, entropic effects will lead to increasingly
shorter chain fragments. From the energy-difference of the linear chain
and the isolated La-dimer, we obtain an estimate for the chain
termination energy of approximately 0.09~eV.  It should be noted, that
experiments often observe shorter chain sequences than predicted from
thermal equilibrium as high-temperature structures are frozen in. 

The electronic structure of the La-chain is analogous to that of the Sr
single chain.\cite{Ashman04} The empty silicon surface has an occupied
and an un-occupied band formed from the dangling bonds of the Si-dimers.
La donates electrons into the upper dangling bond band. Those dangling
bond states, which become filled, are shifted down in energy due to the
change in hybridization on the one side and due to the proximity of the
positive La-cations on the other side.  

\begin{figure}
\includegraphics[angle=0,width=8cm,draft=false,clip=true]{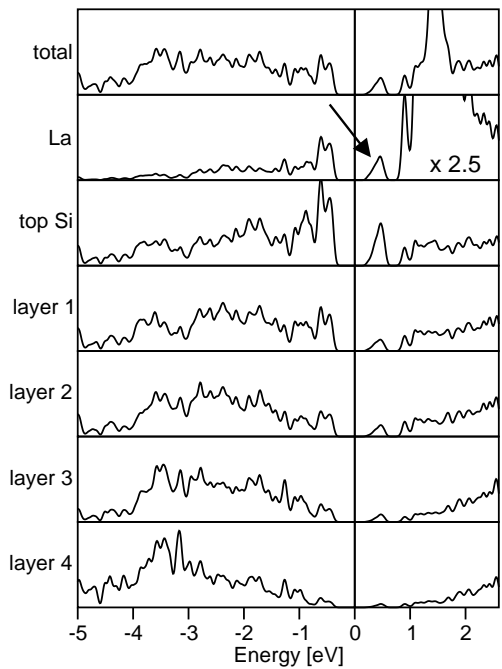}
\caption {Layer resolved density of states of 1/5~ML. The arrow
indicates the upper dangling bond bands in the gap of silicon. The La
panel was magnified by a factor of 2.5. The seemingly large gap of
the silicon substrate is due to finite size effects and also found for
the clean silicon surface at this slab thickness. For bulk silicon we
obtain a "typical" GGA value of 0.65\,eV. This DOS corresponds to
the supercell outlined in Fig.~\ref{fig:lacondchain}\,b.}
\label{fig:5th_dos} 
\end{figure}

\section{Condensed chains}
\label{sec:condchains}

With increasing coverage, the chains become closer packed. In the case
of Sr, there was a preference for a periodicity of $(2n+1)$ surface
lattice spacings along the Si-dimer row direction.\cite{Ashman04} This
restriction has been attributed to the requirement that every cation
be surrounded by four Si-atoms with filled dangling bonds, and that
there is no frustration of the Si-dimer buckling, i.e. adjacent
Si-dimers are buckled antiparallel.

\begin{figure}
\includegraphics[angle=0,width=8cm,draft=false,clip=true]{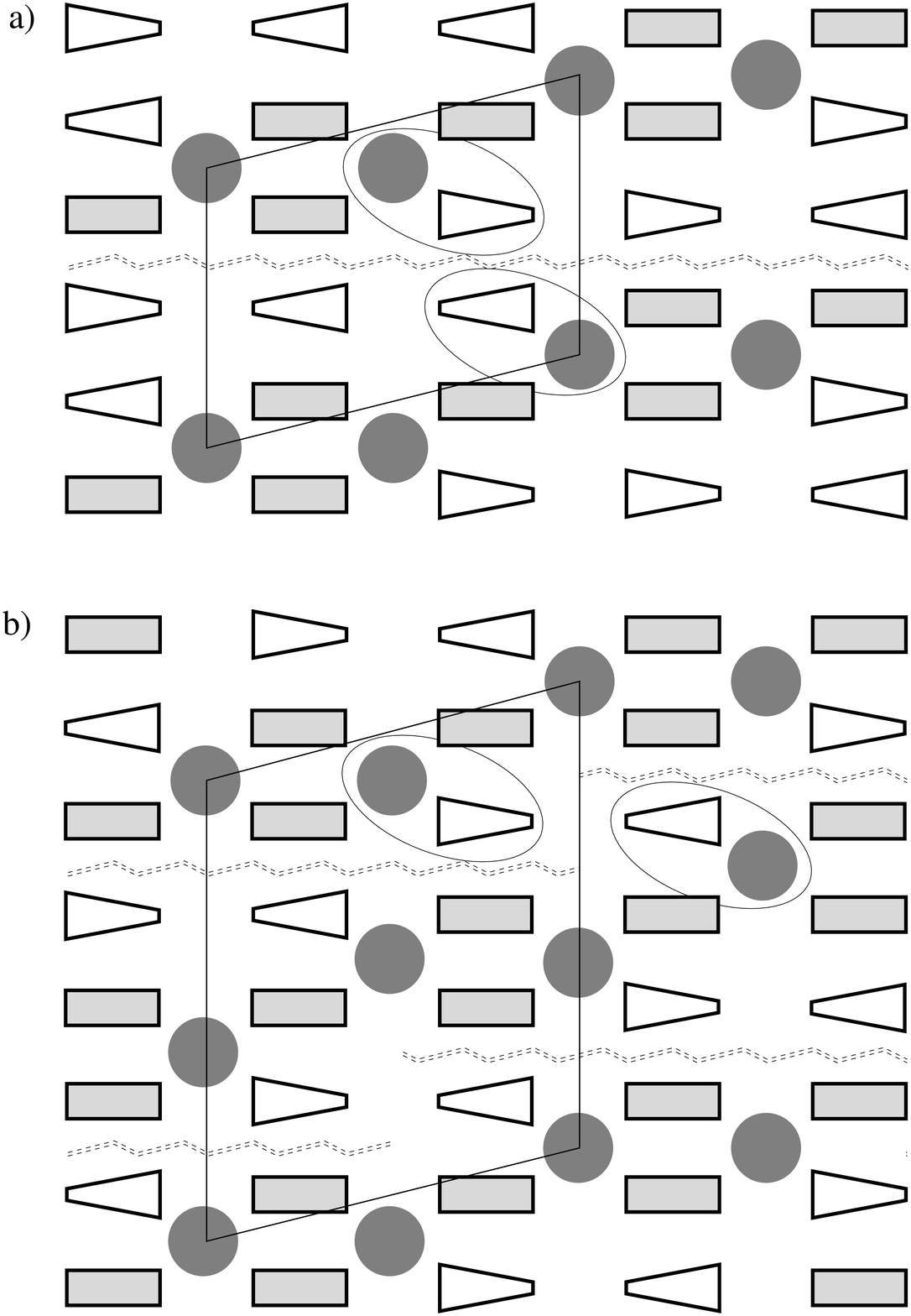}
\caption {Schematic representation of the two packing types of double
stepped La chains at their maximum condensed coverage.  The
reconstruction in panel a) consists of parallel La chains. The chains
in panel b) are also displaced by one valley orthogonal to the Si-dimer
rows.  The ovals indicate the pinning of the Si-dimer buckling by a La
ad-atom.  The dashed double zig-zag lines shows the positions where,
in case of structure a), an even number of Si-dimers can be inserted in
order to arrive at more dilute coverages. In case of structure b) an
arbitrary number of Si-dimers can be inserted, as the buckling of each
row is just pinned on one side. The calculational supercells are
outlined.}
\label{fig:lacondchain} 
\end{figure}

For La the situation is more complex. Due to the longer periodicity of
the La chains compared to those of Sr, there are two families of chain
packing for La as shown in Fig.~\ref{fig:lacondchain}. In the first
family the La chains are displaced only parallel to the Si-dimer row
direction.  In the second family the chains are in addition displaced
perpendicular to the Si-dimer row.

The first family has a preference of $(2n+1)$ surface lattice spacings
along the dimer row as in the case of Sr adsorption. The spacing in
the second family is arbitrary. The reason is that in family one, the
buckling of every second Si-dimer row is pinned on both sides by two
neighboring La chains (see Fig.~\ref{fig:lacondchain}a). A Si-dimer is
pinned, if its buckling is determined by the Coulomb attraction of its
raised, and thus negatively charged, Si atom to a nearby La ion. Since
the buckling alternates along the Si-dimer row, this pinning can lead to
indirect, long-ranged interaction between different La-chains.

In the second family the buckling of every Si-dimer row is pinned only
at one La-chain as seen in Fig.~\ref{fig:lacondchain}b, while there is
no preference of the Si-dimer buckling at the other La-chain.  Thus for
La we find -- in contrast to Sr\cite{Ashman04} -- arbitrary chain
spacings.

The closest packing of La-chains before they merge is 1/5~ML.  We
consider two La-chains merged if La atoms of different La chains occupy
nearest-neighbor $A$ sites within one valley.  We predict a distinct
phase at this coverage as seen in Fig.~\ref{fig:phase_diagram} and
discussed later. This structure, shown in Fig.~\ref{fig:lacondchain}b,
is derived from chains of the second family. An explanation for finding
a phase at 1/5~ML is that the energy at higher coverage increases due to
the electrostatic interaction of the La atoms within one valley.
For the first family, the highest possible coverage before La-chains
merge is 1/6~ML (Fig.~\ref{fig:lacondchain}a).

Note that the chains can change their direction without appreciable
energy cost as shown in Fig.~\ref{fig:lachain}b. Experimentally measured
diffraction patterns would reflect a configurational average.

The layer resolved density of states is shown in Fig.~\ref{fig:5th_dos}.
We see that the Fermi-level lies in a band gap of the surface. Above the
Fermi-level and still in the band-gap of bulk Si, surface bands are
formed, which originate from the remaining empty dangling bonds of the
buckled Si-dimers.  As in the case of Sr, these states form flat bands in the
band-gap of silicon, which approximately remain at their energetic
position as the La coverage is increased. Its density of states,
however, scales with the number of empty dangling bonds. 

\section{The canonical surface at 1/3 ML coverage}
\label{sec:third}

\begin{figure}
\includegraphics[angle=0,width=6.4cm,draft=false,clip=true]{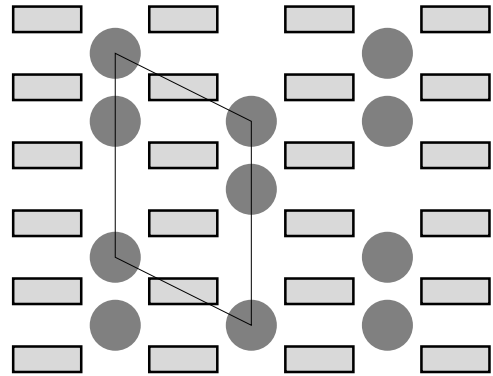}
\caption {Schematic representation of the most stable reconstruction at
the canonical coverage of 1/3~ML. All Si-dimer dangling bonds are filled.
This structure can be thought of as the condensed chain structure in
Fig.~\ref{fig:lacondchain} b) with a reduced chain spacing. The
calculational supercell cell is outlined.}
\label{fig:thirdml} 
\end{figure}

If the spacing of the chains is further reduced, they condense at 1/3
ML to the structure shown in Fig.~\ref{fig:thirdml}.  

There are several versions of this structure type. They have a repeating
sequence of two La-atoms and one vacant $A$ site in each valley in
common. The relative displacement of this sequence from one valley to
the next, however, may differ. We investigated several structures and
found the one shown in Fig.~\ref{fig:thirdml} to be the most stable.

A structure with a sequence of four $A$ sites occupied with metal ions
separated by two empty $A$ sites, has been the most favorable structure
at this coverage in the case of Sr adsorption.\cite{Ashman04}  For La,
however, this configuration is energetically unfavorable.

At a coverage of 1/3 ML, all silicon dangling bonds are filled due to
the electrons provided by the La ad-atoms. This surface is isoelectronic
to the Sr covered surface at 1/2~ML.\cite{Ashman04}  For the Sr-covered
silicon surface, the increased oxidation resistance of the corresponding
1/2 ML structure has been observed experimentally.\cite{Liang01}
Similarly we suggest that the surface covered with 1/3~ML of La will
have an increased oxidation resistance.

\begin{figure}
\includegraphics[angle=0,width=8cm,draft=false,clip=true]{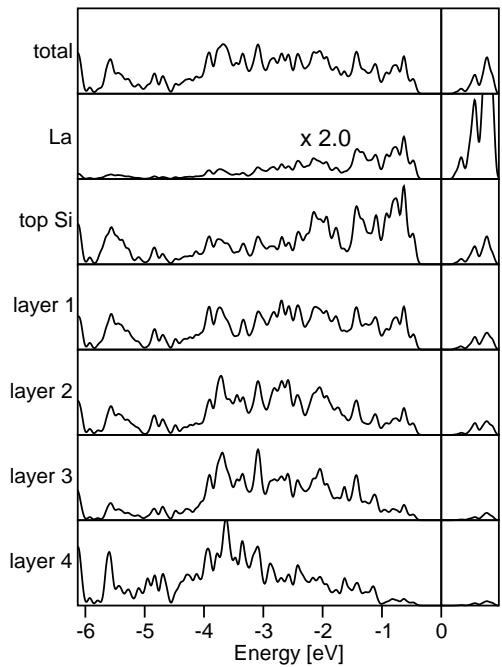}
\caption {Layer resolved density of states of 1/3~ML. The La panel was
magnified by a factor of 2. Compare Fig.~\ref{fig:5th_dos} for a
discussion about the Si band gap. This DOS corresponds to the supercell
outlined in Fig.~\ref{fig:thirdml}.}
\label{fig:3rd_dos} 
\end{figure}

In Fig.~\ref{fig:3rd_dos} we show the layer-resolved density of states
of the most stable structure at 1/3~ML. In analogy to the 1/2~ML covered
Sr surface, there are no surface states deep in the band gap of
silicon, because all Si-dimer dangling bonds are filled and shifted into
the valence band due to the electrostatic attraction of the electrons to
the positive La ions.  Note, however, that in contrast to the canonical
surface coverage of Sr on Si(001) at a coverage of 1/2~ML, the canonical
La surface exhibits vacant $A$-sites.

\section{Transition from La$^{3+}$ to La$^{2+}$ above 1/3 ML}
\label{sec:2+3+}

Up to the canonical coverage of 1/3~ML, all thermodynamically stable
reconstructions could be explained by La being in the $3+$ oxidation
state. In contrast to the isolated La-atoms and La-dimers, the oxidation
state can clearly be identified from the number of unbuckled Si-dimers:
Each unbuckled dimer has received two electrons.  A $3+$ oxidation state
is also consistent with the density of states.

If we follow the picture that emerged from Sr, we would anticipate that
increasing the coverage above 1/3~ML in case of La would lead to filling
the Si-dimer antibonds, which results in a breaking up of the dimer
bonds.  For La the situation is different: the La-$d$ band is located at
much lower energies as compared to Sr.  Therefore the energy to break
the Si-dimer bonds is larger than that to add electrons into the La
$d$-shell.  As a result we find that La changes its oxidation state from
$3+$ to $2+$.  Oxidation states of La that are even lower are
unfavorable due to the Coulomb repulsion of electrons within the La-$d$
and $f$ shells. Thus the structures above 1/3~ML can be explained in
terms of La$^{2+}$ ions and are similar to those found for
Sr.\cite{Ashman04}

\begin{figure}
\includegraphics[angle=0,width=8cm,draft=false,clip=true]{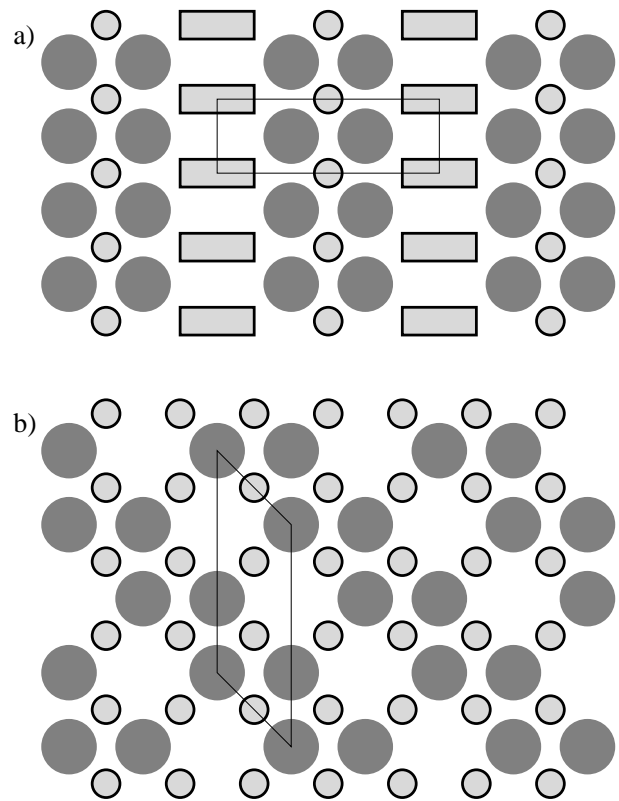}
\caption {The lowest energy structures at 2/3~ML in the 2+ (a) and 3+
(b) regime. The calculational supercells are outlined.} 
\label{fig:2_3rds} 
\end{figure}

It may be instructive to compare two structures with different oxidation
states of La.  A good example is found at a coverage of 2/3~ML: The
lowest energy structure is a $(3\times 1)$ reconstruction already found
for Sr\cite{Ashman04} and depicted in Fig.~\ref{fig:2_3rds}a. This is a
clear 2+ structure.  Since every Si-dimer only accepts two electrons,
they can just accommodate two of the three valence electrons of La. The
lowest structure with formal La$^{3+}$ ions, which can clearly be
identified as having all Si-dimer bonds broken, is shown in
Fig.~\ref{fig:2_3rds}b. It has an energy which is 0.36\,eV per La atom
higher than the structure with La$^{2+}$ ions.

At 1/2 ML, we find a structure where all $A$ sites are occupied to be
most stable.  There the La $d$-states are partially occupied.  We
confirmed that the system is not spin polarized.  

\begin{figure}
\includegraphics[angle=0,width=8cm,draft=false,clip=true]{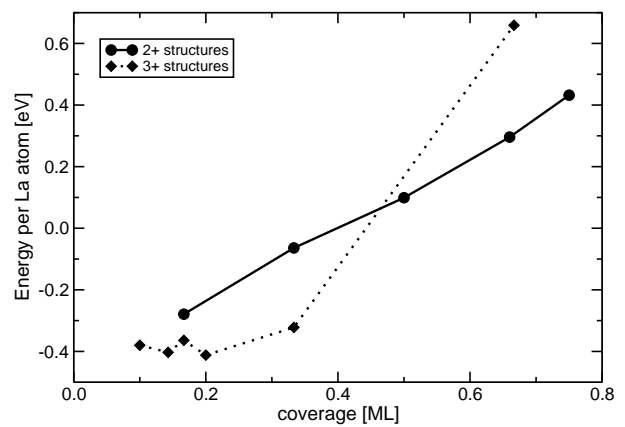}
\caption {Crossover of the total energy surfaces of the 2+ and 3+ regime
beyond the canonical coverage of 1/3~ML.} 
\label{fig:crossover} 
\end{figure}

The crossover of the energy surfaces of the 2+ and the 3+ structures is
shown in Fig.~\ref{fig:crossover} using a set of surface reconstructions,
for which the charge state can be determined unambiguously.  It can be
clearly seen that the 2+ structures become significantly more stable
above 1/2~ML. From Fig.~\ref{fig:enevscov} it is apparent that the
energy rises sharply as the La atoms cross over to the 2+ oxidation
state beyond the canonical interface at a coverage of 1/3~ML.

\section{Phase diagram}
\label{sec:phasediagram}

\begin{figure}

\includegraphics[angle=0,width=8cm,draft=false,clip=true]{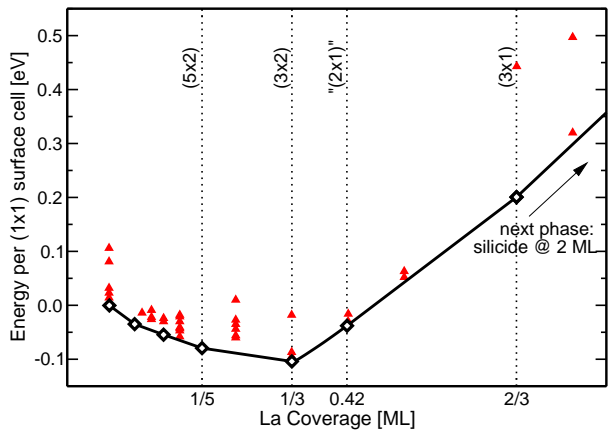}

\caption {(Color online) The surface energy\cite{enevscov} versus
coverage.  
The open diamonds correspond to the thermodynamically accessible surface
structures while the triangles mark metastable structures.
Compare Figs.~\ref{fig:lacondchain}b, \ref{fig:thirdml}
and~\ref{fig:2_3rds}a for the structures at 1/5, 1/3 and 2/3~ML,
respectively.  At 0.42~ML we predict a $(2\times 1)$ reconstruction
which originates from the half-ML structural template with a La vacancy
concentration of 17\,\% (see discussion in the text).} 

\label{fig:enevscov}
\end{figure}

\begin{figure}
\includegraphics[angle=0,width=8cm,draft=false,clip=true]{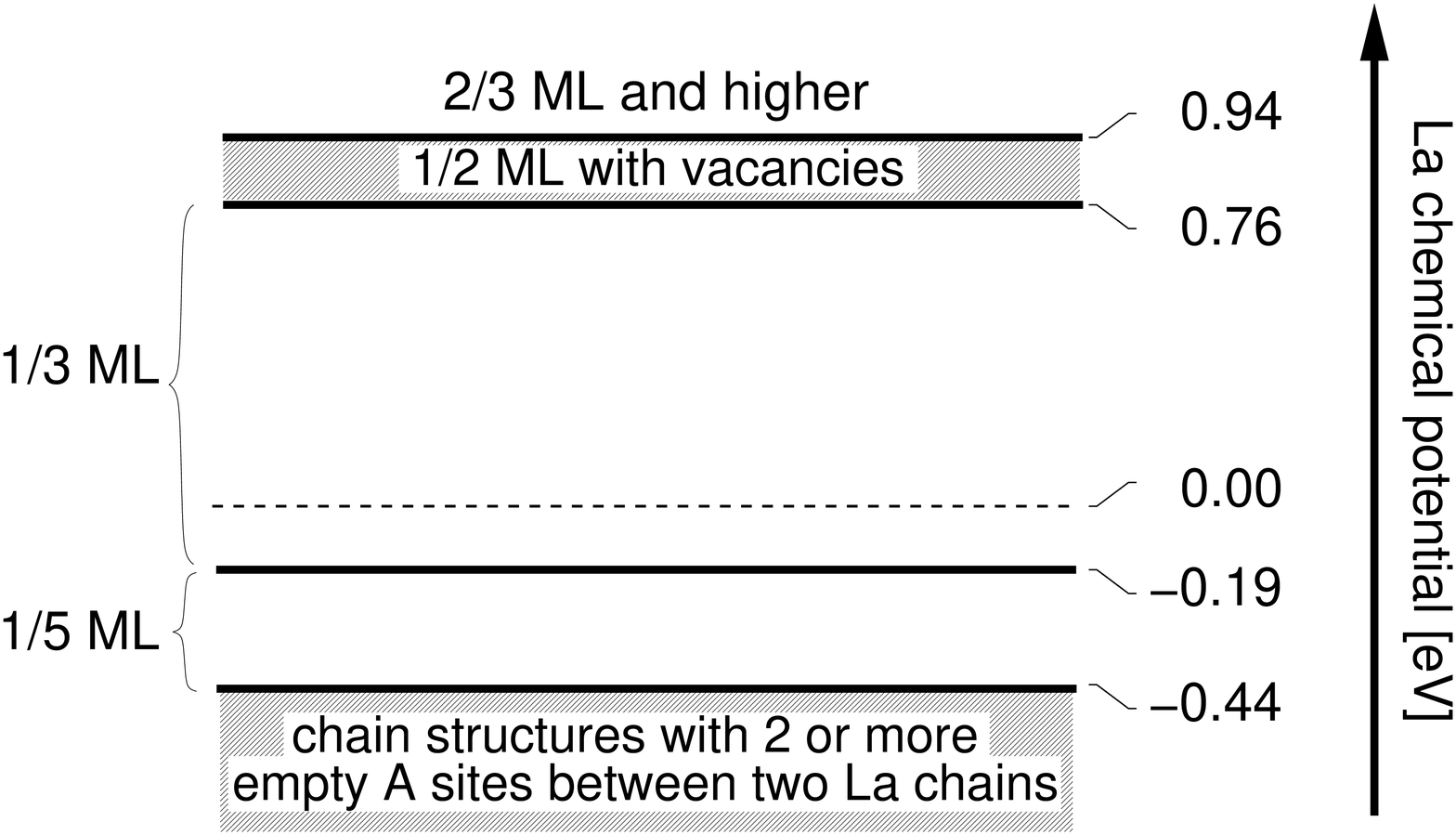}
\caption {The phase diagram for La adsorption on Si(001).}
\label{fig:phase_diagram} 
\end{figure}

Based on the surface energies composed in Fig.~\ref{fig:enevscov} we
extracted the zero-Kelvin phase diagram shown in
Fig.~\ref{fig:phase_diagram}. The slope of the line-segments of the
lower envelope in Fig.~\ref{fig:enevscov} corresponds to the chemical
potential, at which the two neighboring phases coexist (for a more
elaborate discussion, refer to Ref.\cite{Ashman04}). The stable phases
are defined by the coverages where two line segments with different
slopes meet. The zero for the La chemical potential has been chosen as
the value at which LaSi$_2$ and silicon coexist.  Consequently, all
phases in regions of positive chemical potentials are in a regime where
the formation of bulk silicides is thermodynamically favorable. 

Below a chemical potential of $-0.44$\,eV we expect single chain
structures as described in Sec.~\ref{sec:chains}.  At 1/5~ML we predict
a distinct phase since this is the highest possible coverage without La
ad-atoms at nearest neighbor $A$ sites (compare
Fig.~\ref{fig:lacondchain}b). At a chemical potential of -0.19\,eV the
stability region of the 1/3~ML coverage (Fig.~\ref{fig:thirdml}) starts.

The transition from the phase at 1/3 ML to the $2\times1$ reconstructed
surface at 1/2 ML, where all $A$-sites are filled, can be described by a
decrease of La-vacancies (compare Fig.~\ref{fig:thirdml} of this
manuscript and Fig.~9 of Ref.~\cite{Ashman04}).  From this point of
view, the phase at 1/3 ML can be described by an ordered array of
La-vacancies in the 1/2~ML structure.  There is an effective repulsion
between La-vacancies due to the repulsion between La-atoms on
neighboring $A$-sites. We describe the total energy by an empirical
model energy of the form $E(c_V)=E_0+E_f\cdot c_V+\Delta\cdot c_V^2$,
where $c_V$ is the concentration of La vacancies, $E_0$ is the energy of
the structure with all $A$-sites filled (1/2~ML), $E_f$ is the formation
energy of an isolated La-vacancy, and $\Delta$ describes the repulsion
between vacancies.  Coexistence between the two phases would result from
a negative value of $\Delta$. In that case, adding an additional ad-atom
to a phase requires more energy than starting a new phase with the next
higher coverage. Between 1/3 and 1/2~ML, however, $\Delta$ is positive
as filling a portion of vacancies is favorable compared to creating
patches of pure 1/2~ML coverage.

We calculated the energy of an adsorption structure with three La atoms
on neighboring $A$ sites separated by one vacancy within one valley. La
triplets in different valleys have been arranged, so that the distance
between vacancies is maximized in order to minimize the repulsive
energy.  Based on the energies at 1/3 and 1/2~ML as well as at the
intermediate coverage of 3/8~ML just described, we can determine the
three parameters $E_0$, $E_f$ and $\Delta$ to be 0.05, -0.56 and
0.26\,eV, respectively. 

At a certain vacancy concentration of $c_V^0 = 17$\,\% (i.e. a
La-coverage of 0.42~ML)  we find a phase boundary with the next stable
phase at 2/3~ML at a chemical potential of 0.94\,eV. According to our
phase diagram, the pure surface reconstruction at 1/2~ML is never
formed.  The shaded region in Fig.~\ref{fig:phase_diagram} corresponds
to 1/2~ML structural template with variable vacancy concentration.

As seen in the phase diagram shown in Fig.~\ref{fig:phase_diagram} bulk
silicide formation becomes thermodynamically stable within the stability
region of the 1/3~ML coverage.  In a growth experiment we would expect
the formation of bulk silicide grains to be delayed beyond a coverage of
1/3\,ML.  The nucleation of silicide grains may suffer from the large
mismatch between bulk silicide phases and silicon.  This is of
particular importance during the initial stages of nucleation because
the strained interface region occupies most of the volume of the grain.

\begin{figure}
\includegraphics[angle=90,width=6cm,draft=false,clip=true]{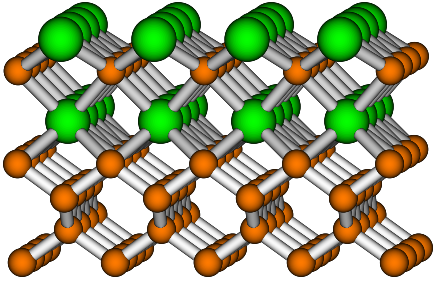}
\caption {(Color online) Silicide overlayer at a coverage of 2~ML.} 
\label{fig:silicide} 
\end{figure}

Thus it may be of interest to know the stability of silicide thin films
on Si(001). We found one such silicide layer which is shown in
Fig.~\ref{fig:silicide}. It consists of a $(1\times1)$ silicon surface
in contact with two La layers that sandwich a layer of Si$^{4-}$ ions in
between.  While we have not performed a thorough search of other
candidates, the energy of this silicide layer indicates that silicide
formation will at the latest be initiated beyond a coverage of 2/3~ML.
We can thus only pin down the onset of silicide formation within a
coverage interval between 1/3 (thermodynamically) and 2/3~ML (including
kinetic considerations).

\section{Conclusions}

In this paper we have investigated the surface structures of La adsorbed on
Si(001) as a function of coverage. We propose a theoretical phase
diagram by relating the phase boundaries at zero temperature to chemical
potentials, which can be converted into partial pressure and temperature
in thermal equilibrium.

Our findings elucidate the chemistry of third row elements on Si(001)
and the phases of La on Si(001), and are expected to provide critical
information for the growth of a wide class of high-k oxides containing
La.  The phase diagram may be used as a guide for the growth of La-based
oxides on Si(001).

\begin{appendix}
\begin{figure}
\centering
\includegraphics[width=8cm]{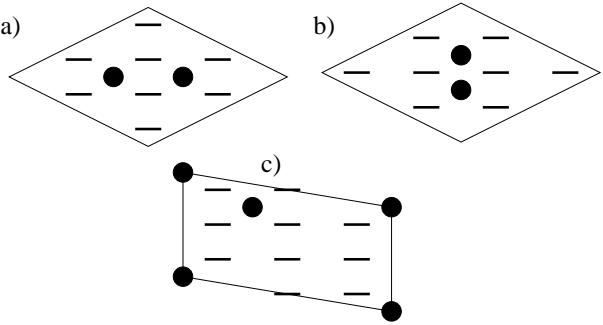}
\caption{Supercells used for the simulation of isolated La 
dimers  shown in Fig.~\ref{fig:dimerorientation}.}
\label{fig:dimersupercells}
\end{figure}

\begin{figure}
\centering
\includegraphics[width=8cm]{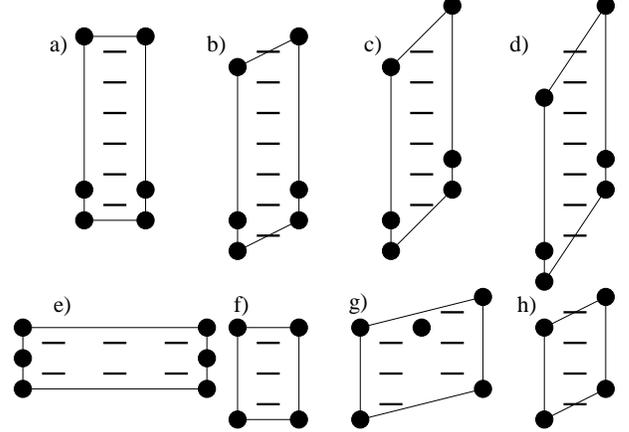}
\caption{Supercells used for the calculation of La chains at 1/6~ML 
as listed in Tab.~\ref{tab:liniearchainenergies}.}
\label{fig:supercells}
\end{figure}

\section{Supercells at dilute coverages}
Fig.~\ref{fig:dimerorientation} shows the three possible La-dimer
orientations on the Si(001) surface.  We did not draw the periodic
images introduced by the calculational supercell in order to emphasize
that fact that this local arrangement corresponds to an isolated dimer.
Fig.~\ref{fig:dimersupercells} sketches the supercells used. They were
chosen in order to avoid frustration of Si dimers due to periodic
images.

Tab.~\ref{tab:liniearchainenergies} summarizes the energetics of chains
structures built from La dimers.  The supercells used in the
corresponding total energy calculations are sketched in
Fig.~\ref{fig:supercells}.

\end{appendix}

\begin{acknowledgments} 

We thank  A.~Dimoulas, J.~Fompeyrine, J.-P.~Loquet, G.~Norga and
C.~Wiemer for useful discussions.  This work has been funded by the
European Commission in the project "INVEST" (Integration of Very High-K
Dielectrics with CMOS Technology) and by the AURORA project of the
Austrian Science Fond.  Parts of the calculations have been performed on
the Computers of the ``Norddeutscher Verbund f\"ur Hoch- und
H\"ochstleistungsrechnen (HLRN)''.  This work has benefited from the
collaborations within the ESF Programme on 'Electronic Structure
Calculations for Elucidating the Complex Atomistic Behavior of Solids
and Surfaces'.

\end{acknowledgments}
%

%

\begin{thebibliography}{99}
\bibitem{spin-pol} The isolated La atom has one unpaired electron. Hence the calculations
have been performed spin polarized. Due to the periodic boundary
conditions, the unpaired electron enters a delocalized, partially
filled band. As a result most of the calculations on isolated defects
produced no net spin.
\bibitem{enevscov} The adsorption energy per $(1\times 1)$ unit cell is
defined as \(E=\big\{\frac{1}{l}\big[E[\mathrm{supercell}]-nE_0[\mathrm{m\
layer\mbox{-}Si\mbox{-}slab}]\big]-E_0[\mathrm{La}] \big\}\cdot X\),
where $E[\mathrm{supercell}]$ is the total energy of the supercell used
for the specific surface reconstruction, $n$ is the number of $(1\times
1)$ surface unit cells contained in that supercell and $m$ is the
slab-thickness in units of silicon layers of the supercell. $l$ denotes
the number of La atoms in the supercell and $X$ the La coverage for that
reconstruction.  This energy can be alternatively calculated using the
energy per La atom multiplied with the coverage $X$. The reference
energies $E_0$ are listed in Tab.~\ref{tab:reference}.
\end{thebibliography}
\end{document}